\documentclass[12pt,twoside,dvips]{article}
\usepackage{amssymb}

\usepackage{amsmath}
\usepackage[spanish,english]{babel}
\usepackage{graphicx}

\begin{document}

\title{Z-Z' mixing in $SU(3)_{c}\otimes SU(3)_{L}\otimes U(1)_{X}$ models
with $\beta$ arbitrary}
\author{Fredy Ochoa$\thanks{%
e-mail: faochoap@unal.edu.co}$ and R. Mart\'{\i}nez$\thanks{%
e-mail: remartinezm@unal.edu.co}$ \and Departamento de F\'{\i}sica,
Universidad Nacional, \\
Bogot\'{a}-Colombia}
\maketitle

\begin{abstract}
We perform a $\chi ^{2}$ fit at 95\% CL to obtain model-dependent bounds to $%
Z_{\mu }-Z_{\mu }^{\prime }$ mixing angle $\theta $ and $Z_{2}$ mass in the
framework of $SU(3)_{c}\otimes SU(3)_{L}\otimes U(1)_{X}$ models with $\beta$
arbitrary. Using experimental results at the $Z$-pole and atomic parity
violation, we obtain allowed regions according to the value of $\beta$ and
depending on the assignment of the quark families in mass eigenstates into
the three different families in weak eigenstates that cancel anomalies.
\end{abstract}

\vspace{-0.3cm}

\vspace{-5mm}

\section{Introduction}

In most of extensions of the SM, new massive and neutral gauge bosons,
called $Z^{\prime}$, are predicted. The phenomenological features that arise
about such boson has been subject of extensive study in the literature \cite%
{zprima}, whose presence is sensitive to experimental observations at low
and high energy, and will be of great interest in the next generation of
colliders (LHC) \cite{godfrey}. In particular, it is possible to study some
phenomenological features associated to this extra neutral gauge boson
through models with gauge symmetry $SU(3)_{c}\otimes SU(3)_{L}\otimes
U(1)_{X},$ also called 331 models. These models arise as an interesting
alternative to explain the origin of generations \cite{Frampton2}, where the three families
are required in order to cancel chiral anomalies completely \cite{anomalias}. An additional
motivation to study these kind of models comes from the fact that they can
also predict the charge quantization for a three family model even when
neutrino masses are added \cite{Pires}.

Although cancellation of anomalies leads to some required conditions \cite%
{fourteen}, such criterion alone still permits an infinite number of 331
models. In these models, the electric charge is defined in general as a
linear combination of the diagonal generators of the group

\begin{equation}
Q=T_{3}+\beta T_{8}+XI.  \label{charge}
\end{equation}

As it has been studied in the literature \cite{fourteen, ten}, the value of
the $\beta $ parameter determines the fermion assignment, and more
specifically, the electric charges of the exotic spectrum. Hence, it is
customary to use this quantum number to classify the different 331 models.
If we want to avoid exotic charges we are led to only two different models
i.e. $\beta =\pm 1/\sqrt{3}$ \cite{fourteen, twelve}. An extensive and
detailed study of models with $\beta$ arbitrary have been carried out in
ref. \cite{331us} for the scalar sector and in ref. \cite{beta-arbitrary}
for the fermionic and gauge sector.

The group structure of these models leads, along with the SM neutral boson $%
Z,$ to the prediction of an additional current associated with the neutral
boson $Z^{\prime }.$ Unlike $Z$-boson whose couplings are family independent
and the weak interactions at low energy are of universal character, the
couplings of $Z^{\prime }$ are different for the three families due to the $%
U(1)_{X}$ values to each of them. Through the $Z-Z^{\prime }$ mixing it is
possible to study the low energy deviations of the $Z$ couplings to the SM
families \cite{ten,Frampton}. In the quark sector each 331-family in the
weak basis can be assigned in three different ways into mass eigenstates. In
this way in a phenomenological analysis, the allowed region associated with
the $Z-Z^{\prime }$ mixing angle and the physical mass $M_{Z_{2}}$ of the
extra neutral boson will depend on the family assignment to the mass states.
This study was carried out in ref. \cite{family-dependence} for the two main
versions of the 331 models corresponding to $\beta =-\sqrt{3} $ \cite{ten}
and $\beta=-\frac{1}{\sqrt{3}}$ \cite{twelve}.

In this work we extend this study to $\beta$ arbitrary through a $\chi ^{2}$
fit at the $Z$-pole to find the allowed region for the mixing angle between
the neutral gauge bosons $Z-Z^{\prime }$, the mass of the $Z_{2}$ boson and
the values of $\beta$ at 95\% CL for three different assignments of the
quark families \cite{Mohapatra}. This analysis will restrict the
possibilities of the infinite number of 331 models, where the values
of the $\beta$ parameter are constrained by phenomenological requirements.
Further, this study will allow us to display restrictions to the main models
with $\beta =-\sqrt{3}$ and $\beta=-\frac{1}{\sqrt{3}}$, which motivate the
exploration of new possibilities of $\beta$.

This paper is organized as follows. Section \ref{sec:331-spectrum} is
devoted to summarize the Fermion, Scalar and Vector boson representations.
In section \ref{sec:neutral-currents} we describe the neutral currents and
the vector and axial vector couplings of the model. In section \ref%
{sec:z-pole} we perform the $\chi ^2$ analysis at the Z-pole including
atomic parity violation at 95\% CL. Finally, section \ref{conclusions}
contains our conclusions.

\section{The 331 spectrum for $\beta$ arbitrary\label{sec:331-spectrum}}

The fermionic spectrum under $SU(3)_{c}\otimes $ $SU(3)_{L}\otimes U(1)_{X}$
is shown in table \ref{tab:espectro} for three families with $\beta$
arbitrary \cite{beta-arbitrary}. We recognize three different possibilities
to assign the physical quarks in each family representation as it is shown
in table \ref{tab:combination}. At low energy, the three models from table %
\ref{tab:combination} are equivalent and there are not any phenomenological
feature that allow us to detect differences between them. In fact, they must
reduce to the SM which is an universal family model in $SU(2)_{L}.$ However,
through the couplings of the three families to the additional neutral
current ($Z^{\prime }$) and the introduction of a mixing angle between $Z$
and $Z^{\prime }$ it is possible to recognize differences among the three
models at the electroweak scale. It is noted that although we write the
spectrum in the weak basis in table \ref{tab:espectro}, we can consider
three realizations in the mass basis in table \ref{tab:combination}.

\begin{table}[tbp]
\begin{center}
\begin{equation*}
\begin{tabular}{||c||c||c||}
\hline\hline
$representation$ & $Q_\psi $ & $X_\psi $ \\ \hline\hline
$\ 
\begin{tabular}{c}
$q_{m^{*}L}=\left( 
\begin{array}{c}
d,s \\ 
-u,-c \\ 
J_1,J_2%
\end{array}
\right) _L\mathbf{3}^{*}$ \\ 
\\ 
\\ 
$d_{m^{*}R}=d_R,s_R:\mathbf{1}$ \\ 
$u_{m^{*}R}=u_R,c_R:\mathbf{1}$ \\ 
$J_{m^{*}R}=J_{1R},J_{2R}:\mathbf{1}$%
\end{tabular}
$ & 
\begin{tabular}{c}
$\left( 
\begin{array}{c}
-\frac 13 \\ 
\frac 23 \\ 
\frac 16+\frac{\sqrt{3}\beta }2%
\end{array}
\right) $ \\ 
\\ 
$-\frac 13$ \\ 
$\frac 23$ \\ 
$\frac 16+\frac{\sqrt{3}}2\beta $%
\end{tabular}
& 
\begin{tabular}{c}
\\ 
$X_{q^{(m)}}^L=-\frac 16-\frac \beta {2\sqrt{3}}$ \\ 
\\ 
\\ 
$X_{u^{(m)}}^R=-\frac 13$ \\ 
$X_{d^{(m)}}^R=\frac 23$ \\ 
$X_{J^{(m)}}^R=\frac 16+\frac{\sqrt{3}}2\beta $%
\end{tabular}
\\ \hline\hline
\begin{tabular}{c}
$q_{3L}=\left( 
\begin{array}{c}
t \\ 
b \\ 
J_3%
\end{array}
\right) _L:\mathbf{3}$ \\ 
\\ 
$u_{3R}=b_R:\mathbf{1}$ \\ 
$d_{3R}=t_R:\mathbf{1}$ \\ 
$J_{3R}=J_{3R}:\mathbf{1}$%
\end{tabular}
& 
\begin{tabular}{c}
$\left( 
\begin{array}{c}
\frac 23 \\ 
-\frac 13 \\ 
\frac 16-\frac{\sqrt{3}\beta }2%
\end{array}
\right) $ \\ 
\\ 
$-\frac 13$ \\ 
$\frac 23$ \\ 
$\frac 16-\frac{\sqrt{3}\beta }2$%
\end{tabular}
& 
\begin{tabular}{c}
\\ 
$X_{q^{(3)}}^L=\frac 16-\frac \beta {2\sqrt{3}}$ \\ 
\\ 
\\ 
$X_b^R=-\frac 13$ \\ 
$X_t^R=\frac 23$ \\ 
$X_{J_3}^R=\frac 16-\frac{\sqrt{3}\beta }2$%
\end{tabular}
\\ \hline\hline
\begin{tabular}{c}
$\ell _{jL}=\left( 
\begin{array}{c}
\nu _e,\nu _\mu ,\nu _\tau \\ 
e^{-},\mu ^{-},\tau ^{-} \\ 
E_1^{-Q_1},E_2^{-Q_1},E_3^{-Q_1}%
\end{array}
\right) _L:\mathbf{3}$ \\ 
\\ 
$\left( e_j^{-}\right) _R=e^{-},\mu ^{-},\tau _R^{-}:\mathbf{1}$ \\ 
$E_j^{-Q_1}=E_1^{-Q_1},E_2^{-Q_1},E_3^{-Q_1}:\mathbf{1}$%
\end{tabular}
& 
\begin{tabular}{c}
$\left( 
\begin{array}{c}
0 \\ 
-1 \\ 
-\frac 12-\frac{\sqrt{3}\beta }2%
\end{array}
\right) $ \\ 
\\ 
$-1$ \\ 
$-\frac 12-\frac{\sqrt{3}\beta }2$%
\end{tabular}
& 
\begin{tabular}{c}
\\ 
$X_{\ell ^{(m)}}^L=-\frac 12-\frac \beta {2\sqrt{3}}$ \\ 
\\ 
\\ 
$X_{e^{(m)}}^R=-1$ \\ 
$X_{E_m}^R=-\frac 12-\frac{\sqrt{3}\beta }2$%
\end{tabular}
\\ \hline\hline
\end{tabular}%
\end{equation*}%
\end{center}
\caption{\textit{Fermionic content for three generations with\ }$\protect%
\beta \ $\textit{arbitrary.}}
\label{tab:espectro}
\end{table}

For the scalar sector, we introduce the triplet field $\chi $ with Vacuum
Expectation Value (VEV) $\left\langle \chi \right\rangle ^{T}=\left( 0,0,\nu
_{\chi }\right) $, which induces the masses to the third fermionic
components. In the second transition it is necessary to introduce two
triplets$\;\rho $ and $\eta $ with VEV $\left\langle \rho \right\rangle
^{T}=\left( 0,\nu _{\rho },0\right) $ and $\left\langle \eta \right\rangle
^{T}=\left( \nu _{\eta },0,0\right) $ in order to give masses to the quarks
of type up and down respectively.

\begin{table}[tbp]
\begin{equation*}
\begin{tabular}{||c||c||c||}
\hline\hline
Representation $A$ & Representation $B$ & Representation $C$ \\ \hline\hline
$%
\begin{tabular}{c}
$q_{mL}=\left( 
\begin{array}{c}
d,s \\ 
-u,-c \\ 
J_{1},J_{2}%
\end{array}%
\right) _{L}:\mathbf{3}^{\ast }$ \\ 
$q_{3L}=\left( 
\begin{array}{c}
t \\ 
b \\ 
J_{3}%
\end{array}%
\right) _{L}:\mathbf{3}$%
\end{tabular}%
\ $ & $%
\begin{tabular}{c}
$q_{mL}=\left( 
\begin{array}{c}
d,b \\ 
-u,-t \\ 
J_{1},J_{3}%
\end{array}%
\right) _{L}:\mathbf{3}^{\ast }$ \\ 
$q_{3L}=\left( 
\begin{array}{c}
c \\ 
s \\ 
J_{2}%
\end{array}%
\right) _{L}:\mathbf{3}$%
\end{tabular}%
\ $ & $%
\begin{tabular}{c}
$q_{mL}=\left( 
\begin{array}{c}
s,b \\ 
-c,-t \\ 
J_{2},J_{3}%
\end{array}%
\right) _{L}:\mathbf{3}^{\ast }$ \\ 
$q_{3L}=\left( 
\begin{array}{c}
u \\ 
d \\ 
J_{1}%
\end{array}%
\right) _{L}:\mathbf{3}$%
\end{tabular}%
\ $ \\ \hline\hline
\end{tabular}%
\end{equation*}%
\caption{\textit{Three different assignments for the $SU(3)_L$ family
representation of quarks}}
\label{tab:combination}
\end{table}

In the gauge boson spectrum associated with the group $SU(3)_{L}\otimes
U(1)_{X},$ we are just interested in the physical neutral sector that
corresponds to the photon, $Z$ and $Z^{\prime },$ which are written in terms
of the electroweak basis for $\beta $ arbitrary as \cite{beta-arbitrary}

\begin{eqnarray}
A_{\mu } &=&S_{W}W_{\mu }^{3}+C_{W}\left( \beta T_{W}W_{\mu }^{8}+\sqrt{%
1-\beta ^{2}T_{W}^{2}}B_{\mu }\right) ,  \notag \\
Z_{\mu } &=&C_{W}W_{\mu }^{3}-S_{W}\left( \beta T_{W}W_{\mu }^{8}+\sqrt{%
1-\beta ^{2}T_{W}^{2}}B_{\mu }\right) ,  \notag \\
Z_{\mu }^{\prime } &=&-\sqrt{1-\beta ^{2}T_{W}^{2}}W_{\mu }^{8}+\beta
T_{W}B_{\mu },
\end{eqnarray}

\noindent where the Weinberg angle is defined as

\begin{equation}
S_{W}=\sin \theta _{W}=\frac{g^{\prime }}{\sqrt{g^{2}+\left( 1+\beta
^{2}\right) g^{\prime 2}}},\quad T_{W}=\tan \theta _{W}=\frac{g^{\prime }}{%
\sqrt{g^{2}+\beta^{2} g^{\prime 2}}}
\end{equation}

\noindent and $g,$ $g^{\prime }$ correspond to the coupling constants of the groups $%
SU(3)_{L}$ and $U(1)_{X}$ respectively. Further, a small mixing angle
between the two neutral currents $Z_{\mu }$ and $Z_{\mu }^{\prime }$ appears
with the following mass eigenstates \cite{beta-arbitrary}

\begin{eqnarray}
Z_{1\mu } &=&Z_{\mu }C_{\theta }+Z_{\mu }^{\prime }S_{\theta };\quad Z_{2\mu
}=-Z_{\mu }S_{\theta }+Z_{\mu }^{\prime }C_{\theta };  \notag \\
\tan \theta &=&\frac{1}{\Lambda +\sqrt{\Lambda ^{2}+1}};\quad \Lambda =\frac{%
-2S_{W}C_{W}^{2}g^{\prime 2}\nu _{\chi }^{2}+\frac{3}{2}S_{W}T_{W}^{2}g^{2}%
\left( \nu _{\eta }^{2}+\nu _{\rho }^{2}\right) }{gg^{\prime }T_{W}^{2}\left[
3\beta S_{W}^{2}\left( \nu _{\eta }^{2}+\nu _{\rho }^{2}\right)
+C_{W}^{2}\left( \nu _{\eta }^{2}-\nu _{\rho }^{2}\right) \right] }.
\label{mix}
\end{eqnarray}

\section{Neutral currents\label{sec:neutral-currents}}

Using the fermionic content from table \ref{tab:espectro}, we obtain the
neutral coupling for the SM fermions \cite{beta-arbitrary}

\begin{eqnarray}
\mathcal{L}^{NC} &=&\sum_{j=1}^{3}\left\{ \frac{g}{2C_{W}}\overline{\text{Q}%
_{j}}\gamma _{\mu }\left[ 2T_{3}P_{L}-2Q_{\text{Q}_{j}}S_{W}^{2}\right] 
\text{Q}_{j}Z^{\mu }\right.  \notag \\
&+&\frac{g}{2C_{W}}\overline{\ell _{j}}\gamma _{\mu }\left[
2T_{3}P_{L}-2Q_{\ell _{j}}S_{W}^{2}\right] \ell _{j}Z^{\mu }  \notag \\
&+&\left. \frac{g^{\prime }}{2T_{W}}\overline{\ell _{j}}\gamma _{\mu }\left[
\left( -2T_{8}-\beta T_{W}^{2}\Lambda _{3}\right) P_{L}+2\beta Q_{\ell
_{j}}T_{W}^{2}P_{R}\right] \ell _{j}Z^{\mu \prime }\right\}  \notag \\
&+&\sum_{m=1}^{2}\frac{g^{\prime }}{2T_{W}}\overline{q_{m}}\gamma _{\mu }%
\left[ \left( 2T_{8}+\beta Q_{q_{m}}T_{W}^{2}\Lambda _{1}\right)
P_{L}+2\beta Q_{q_{m}}T_{W}^{2}P_{R}\right] q_{m}Z^{\mu \prime }  \notag \\
&+&\frac{g^{\prime }}{2T_{W}}\overline{q_{3}}\gamma _{\mu }\left[ \left(
-2T_{8}+\beta Q_{q_{3}}T_{W}^{2}\Lambda _{2}\right) P_{L}+2\beta
Q_{q_{3}}T_{W}^{2}P_{R}\right] q_{3}Z^{\mu \prime },  \label{lag-1}
\end{eqnarray}

where Q$_{j}$ with $j=1,2,3$ has been written in a SM-like notation i.e. it
refers to triplets of quarks associated with the three generations of quarks
(SM does not make difference in the family representations). On the other
hand, the coupling of the exotic gauge boson ($Z_{\mu }^{\prime }$) with the
two former families are different from the ones involving the third family.
This is because the third familiy transforms differently as it was remarked
in table \ref{tab:espectro}. Consequently, there are terms where only the
components $m=1,2$ are summed, leaving the third one in a term apart. $%
Q_{q_{j}}$ are the electric charges. The Gell-Mann matrices $T_{3}$ $=\frac{1%
}{2}diag(1,-1,0)$ and $T_{8}=\frac{1}{2\sqrt{3}}diag(1,1,-2)$ are introduced
in the notation. We also define $\Lambda _{1}=diag(-1,\frac{1}{2},2)$, $%
\Lambda _{2}=diag(\frac{1}{2},-1,2)$ and the projectors $P_{R,L}=\frac{1}{2}%
(1\pm \gamma _{5}).$ Finally, $\ell _{j}$ denote the leptonic triplets with $%
Q_{\ell _{j}}$ denoting their electric charges and $\Lambda
_{3}=diag(1,1,2Q_{1})$ with $Q_{1}$ defined as the electric charge of the
exotic leptons $E_{j}$ in table \ref{tab:espectro}. Following the same
procedure as ref. \cite{family-dependence}, the neutral lagrangian (\ref%
{lag-1}) can be written as

\begin{eqnarray}
\mathcal{L}^{NC} &=&\sum_{j=1}^{3}\left\{ \frac{g}{2C_{W}}\overline{\text{Q}%
_{j}}\gamma _{\mu }\left[ G_{V}^{Q_{j}}-G_{A}^{Q_{j}}\gamma _{5}\right] 
\text{Q}_{j}Z_{1}^{\mu }+\frac{g}{2C_{W}}\overline{\ell _{j}}\gamma _{\mu }%
\left[ G_{V}^{\ell _{j}}-G_{A}^{\ell _{j}}\gamma _{5}\right] \ell
_{j}Z_{1}^{\mu }\right.  \notag \\
&&+\left. \frac{g}{2C_{W}}\overline{\text{Q}_{j}}\gamma _{\mu }\left[ 
\overset{\sim }{G}_{V}^{Q_{j}}-\overset{\sim }{G}_{A}^{Q_{j}}\gamma _{5}%
\right] \text{Q}_{j}Z_{2}^{\mu }+\frac{g}{2C_{W}}\overline{\ell _{j}}\gamma
_{\mu }\left[ \overset{\sim }{G}_{V}^{\ell _{j}}-\overset{\sim }{G}%
_{A}^{\ell _{j}}\gamma _{5}\right] \ell _{j}Z_{2}^{\mu }\right\} ,
\label{lag-3}
\end{eqnarray}

\noindent where the couplings associated with $Z_{1\mu }$ are

\begin{eqnarray}
G_{V,A}^{f} &=&g_{V,A}^{f}+\delta g_{V,A}^{f},  \notag  \label{coupling-2} \\
\delta g_{V,A}^{f} &=&\overset{\sim }{g}_{V,A}^{f}S_{\theta },  \label{coup2}
\end{eqnarray}

\noindent the couplings associated with $Z_{2\mu }$ are

\begin{eqnarray}
\overset{\sim }{G}_{V,A}^{f} &=&\overset{\sim }{g}_{V,A}^{f}-\delta \overset{%
\sim }{g}_{V,A}^{f},  \notag \\
\delta \overset{\sim }{g}_{V,A}^{f} &=&g_{V,A}^{f}S_{\theta }.  \label{coup3}
\end{eqnarray}

\noindent and the vector and axial vector couplings are given by

\begin{eqnarray}
g_{V}^{f} &=&T_{3}-2Q_{f}S_{W}^{2},\qquad g_{A}^{f}=T_{3}  \notag \\
\overset{\sim }{g}_{V,A}^{q_{m}} &=&\frac{C_{W}^{2}}{\sqrt{1-\left( 1+\beta
^{2}\right) S_{W}^{2}}}\left[ T_{8}+\beta Q_{q_{m}}T_{W}^{2}\left( \frac{1}{2%
}\Lambda _{1}\pm 1\right) \right]  \notag \\
\overset{\sim }{g}_{V,A}^{q_{3}} &=&\frac{C_{W}^{2}}{\sqrt{1-\left( 1+\beta
^{2}\right) S_{W}^{2}}}\left[ -T_{8}+\beta Q_{q_{3}}T_{W}^{2}\left( \frac{1}{%
2}\Lambda _{2}\pm 1\right) \right]  \notag \\
\overset{\sim }{g}_{V,A}^{\ell _{j}} &=&\frac{C_{W}^{2}}{\sqrt{1-\left(
1+\beta ^{2}\right) S_{W}^{2}}}\left[ -T_{8}-\beta T_{W}^{2}\left( \frac{1}{2%
}\Lambda _{3}\mp Q_{\ell _{j}}\right) \right] .  \label{coup1}
\end{eqnarray}

In the above equations we took into account the small mixing angle given by
eq. (\ref{mix}), where we did $C_{\theta}\simeq1$

\section{Z-Pole Observables\label{sec:z-pole}}

The couplings of the $Z_{1\mu }$ in eq. (\ref{lag-3}) have the same form as
the SM neutral couplings but by replacing the vector and axial vector
couplings $g_{V,A}^{SM}$ by $G_{V,A}=g_{V,A}^{SM}+\delta g_{V,A},$ where $%
\delta g_{V,A}$ (given by eq. (\ref{coup2})) is a correction due to the
small $Z_{\mu }-Z_{\mu }^{\prime }$ mixing angle $\theta .$ For this reason
all the analytical parameters at the Z pole have the same SM-form but with
small correction factors that depend on the family assignment. The partial
decay widths of $Z_{1}$ into fermions $f\overline{f}$ is described by \cite%
{one, pitch}:

\begin{equation}
\Gamma _{f}^{SM}=\frac{N_{c}^{f}G_{f}M_{Z_{1}}^{3}}{6\sqrt{2}\pi }\rho _{f}%
\left[ \frac{3\beta _{K}-\beta _{K}^{3}}{2}\left( g_{V}^{f}\right)
^{2}+\beta _{K}^{3}\left( g_{A}^{f}\right) ^{2}\right] R_{QED}R_{QCD},
\label{partial-decay}
\end{equation}

\noindent where $N_{c}^{f}=1$, 3 for leptons and quarks respectively, $%
R_{QED,QCD}$ are global final-state QED and QCD corrections, and $\beta _{K}=%
\sqrt{1-\frac{4m_{b}^{2}}{M_{Z}^{2}}}$ considers kinematic corrections only
important for the $b$-quark. Universal electroweak corrections sensitive to
the top quark mass are taken into account in $\rho _{f}=1+\rho _{t}$ and in $%
g_{V}^{SM}$ which is written in terms of an effective Weinberg angle \cite%
{one}

\begin{equation}
\overline{S_{W}}^{2}=\kappa _{f}S_{W}^{2}=\left( 1+\frac{\rho _{t}}{T_{W}^{2}%
}\right) S_{W}^{2},  \label{effective-angle}
\end{equation}

\noindent with $\rho _{t}=3G_{f}m_{t}^{2}/8\sqrt{2}\pi ^{2}$. Non-universal
vertex corrections are also taken into account in the $Z_{1}\overline{b}b$
vertex with additional one-loop leading terms given by \cite{one, pitch}

\begin{equation}
\rho _{b}\rightarrow \rho _{b}-\frac{4}{3}\rho _{t}\text{ and }\kappa
_{b}\rightarrow \kappa _{b}+\frac{2}{3}\rho _{t}.  \label{Zb-vertex}
\end{equation}

Table \ref{tab:observables} resumes some observables, with their
experimental values from CERN collider (LEP), SLAC Liner Collider (SLC) and
data from atomic parity violation \cite{one}, the SM predictions and the
expressions predicted by 331 models. We use $M_{Z_{1}}=91.1876$ $GeV,$ $%
m_{t}=176.9$ $GeV$, $S_{W}^{2}=0.2314$, and for $m_{b}$ we use \cite{greub}

\begin{equation*}
\overline{m}_{b}(\mu \rightarrow M_{Z_{1}})=m_{b}\left[ 1+\frac{\alpha
_{S}(\mu )}{\pi }\left( \ln \frac{m_{b}^{2}}{\mu ^{2}}-\frac{4}{3}\right) %
\right] ,
\end{equation*}

\noindent with $m_{b}\approx 4.5$ $GeV$ the pole mass, $\overline{m}_{b}(\mu
\rightarrow M_{Z_{1}})$ the running mass at $M_{Z_{1}}$ scale in the $%
\overline{MS}$ scheme, and $\alpha _{S}(M_{Z_{1}})=0.1213\pm 0.0018$ the
strong coupling constant.

The 331 predictions from table \ref{tab:observables} are expressed in terms
of SM values corrected by

\begin{eqnarray}
\delta _{Z} &=&\frac{\Gamma _{u}^{SM}}{\Gamma _{Z}^{SM}}(\delta _{u}+\delta
_{c})+\frac{\Gamma _{d}^{SM}}{\Gamma _{Z}^{SM}}(\delta _{d}+\delta _{s})+%
\frac{\Gamma _{b}^{SM}}{\Gamma _{Z}^{SM}}\delta _{b}+3\frac{\Gamma _{\nu
}^{SM}}{\Gamma _{Z}^{SM}}\delta _{\nu }+3\frac{\Gamma _{e}^{SM}}{\Gamma
_{Z}^{SM}}\delta _{\ell };  \notag \\
\delta _{had} &=&R_{c}^{SM}(\delta _{u}+\delta _{c})+R_{b}^{SM}\delta _{b}+%
\frac{\Gamma _{d}^{SM}}{\Gamma _{had}^{SM}}(\delta _{d}+\delta _{s});  \notag
\\
\delta _{\sigma } &=&\delta _{had}+\delta _{\ell }-2\delta _{Z};  \notag \\
\delta A_{f} &=&\frac{\delta g_{V}^{f}}{g_{V}^{f}}+\frac{\delta g_{A}^{f}}{%
g_{A}^{f}}-\delta _{f},  \label{shift1}
\end{eqnarray}

\noindent where for the light fermions

\begin{equation}
\delta _{f}=\frac{2g_{V}^{f}\delta g_{V}^{f}+2g_{A}^{f}\delta g_{A}^{f}}{%
\left( g_{V}^{f}\right) ^{2}+\left( g_{A}^{f}\right) ^{2}},  \label{shift2}
\end{equation}

\noindent while for the $b$-quark

\begin{equation}
\delta _{b}=\frac{\left( 3-\beta _{K}^{2}\right) g_{V}^{b}\delta
g_{V}^{b}+2\beta _{K}^{2}g_{A}^{b}\delta g_{A}^{b}}{\left( \frac{3-\beta
_{K}^{2}}{2}\right) \left( g_{V}^{b}\right) ^{2}+\beta _{K}^{2}\left(
g_{A}^{b}\right) ^{2}}.  \label{shift3}
\end{equation}

\noindent The above expressions are evaluated in terms of the effective
Weinberg angle from eq. (\ref{effective-angle}). For the predicted SM partial decay given by eq. (\ref{partial-decay}), we use
the values from ref. \cite{one}

\begin{table}[tbp]
\begin{center}
$%
\begin{tabular}{|c|c|c|c|}
\hline
Quantity & Experimental Values & Standard Model & 331 Model \\ \hline
$\Gamma _{Z}$ $\left[ GeV\right] $ & 2.4952 $\pm $ 0.0023 & 2.4972 $\pm $
0.0012 & $\Gamma _{Z}^{SM}\left( 1+\delta _{Z}\right) $ \\ \hline
$\Gamma _{had}$ $\left[ GeV\right] $ & 1.7444 $\pm $ 0.0020 & 1.7435 $\pm $
0.0011 & $\Gamma _{had}^{SM}\left( 1+\delta _{had}\right) $ \\ \hline
$\Gamma _{\left( \ell ^{+}\ell ^{-}\right) }$ $MeV$ & 83.984 $\pm $ 0.086 & 
84.024 $\pm $ 0.025 & $\Gamma _{\left( \ell ^{+}\ell ^{-}\right)
}^{SM}\left( 1+\delta _{\ell }\right) $ \\ \hline
$\sigma _{had}$ $\left[ nb\right] $ & 41.541 $\pm $ 0.037 & 41.472 $\pm $
0.009 & $\sigma _{had}^{SM}\left( 1+\delta _{\sigma }\right) $ \\ \hline
$R_{e}$ & 20.804 $\pm $ 0.050 & 20.750 $\pm $ 0.012 & $R_{e}^{SM}\left(
1+\delta _{had}+\delta _{e}\right) $ \\ \hline
$R_{\mu }$ & 20.785 $\pm $ 0.033 & 20.751 $\pm $ 0.012 & $R_{\mu
}^{SM}\left( 1+\delta _{had}+\delta _{\mu }\right) $ \\ \hline
$R_{\tau }$ & 20.764 $\pm $ 0.045 & 20.790 $\pm $ 0.018 & $R_{\tau
}^{SM}\left( 1+\delta _{had}+\delta _{\tau }\right) $ \\ \hline
$R_{b}$ & 0.21638 $\pm $ 0.00066 & 0.21564 $\pm $ 0.00014 & $%
R_{b}^{SM}\left( 1+\delta _{b}-\delta _{had}\right) $ \\ \hline
$R_{c}$ & 0.1720 $\pm $ 0.0030 & 0.17233 $\pm $ 0.00005 & $R_{c}^{SM}\left(
1+\delta _{c}-\delta _{had}\right) $ \\ \hline
$A_{e}$ & 0.15138 $\pm $ 0.00216 & 0.1472 $\pm $ 0.0011 & $A_{e}^{SM}\left(
1+\delta A_{e}\right) $ \\ \hline
$A_{\mu }$ & 0.142 $\pm $ 0.015 & 0.1472 $\pm $ 0.0011 & $A_{\mu
}^{SM}\left( 1+\delta A_{\mu }\right) $ \\ \hline
$A_{\tau }$ & 0.136 $\pm $ 0.015 & 0.1472 $\pm $ 0.0011 & $A_{\tau
}^{SM}\left( 1+\delta A_{\tau }\right) $ \\ \hline
$A_{b}$ & 0.925 $\pm $ 0.020 & 0.9347 $\pm $ 0.0001 & $A_{b}^{SM}\left(
1+\delta A_{b}\right) $ \\ \hline
$A_{c}$ & 0.670 $\pm $ 0.026 & 0.6678 $\pm $ 0.0005 & $A_{c}^{SM}\left(
1+\delta A_{c}\right) $ \\ \hline
$A_{s}$ & 0.895 $\pm $ 0.091 & 0.9357 $\pm $ 0.0001 & $A_{s}^{SM}\left(
1+\delta A_{s}\right) $ \\ \hline
$A_{FB}^{\left( 0,e\right) }$ & 0.0145 $\pm $ 0.0025 & 0.01626 $\pm $ 0.00025
& $A_{FB}^{(0,e)SM}\left( 1+2\delta A_{e}\right) $ \\ \hline
$A_{FB}^{\left( 0,\mu \right) }$ & 0.0169 $\pm $ 0.0013 & 0.01626 $\pm $
0.00025 & $A_{FB}^{(0,\mu )SM}\left( 1+\delta A_{e}+\delta A_{\mu }\right) $
\\ \hline
$A_{FB}^{\left( 0,\tau \right) }$ & 0.0188 $\pm $ 0.0017 & 0.01626 $\pm $
0.00025 & $A_{FB}^{(0,\tau )SM}\left( 1+\delta A_{e}+\delta A_{\tau }\right) 
$ \\ \hline
$A_{FB}^{\left( 0,b\right) }$ & 0.0997 $\pm $ 0.0016 & 0.1032 $\pm $ 0.0008
& $A_{FB}^{(0,b)SM}\left( 1+\delta A_{e}+\delta A_{b}\right) $ \\ \hline
$A_{FB}^{\left( 0,c\right) }$ & 0.0706 $\pm $ 0.0035 & 0.0738 $\pm $ 0.0006
& $A_{FB}^{(0,c)SM}\left( 1+\delta A_{e}+\delta A_{c}\right) $ \\ \hline
$A_{FB}^{\left( 0,s\right) }$ & 0.0976 $\pm $ 0.0114 & 0.1033 $\pm $ 0.0008
& $A_{FB}^{(0,s)SM}\left( 1+\delta A_{e}+\delta A_{s}\right) $ \\ \hline
$Q_{W}(Cs)$ & $-$72.69 $\pm $ 0.48 & $-$73.19 $\pm $ 0.03 & $%
Q_{W}^{SM}\left( 1+\delta Q_{W}\right) $ \\ \hline
\end{tabular}%
\ \ $%
\end{center}
\caption{\textit{The parameters for experimental values, SM predictions and
331 corrections. The values are taken from ref. \protect\cite{one}}}
\label{tab:observables}
\end{table}

The weak charge is written as

\begin{equation}
Q_{W}=Q_{W}^{SM}+\Delta Q_{W}=Q_{W}^{SM}\left( 1+\delta Q_{W}\right) ,
\label{weak}
\end{equation}
where $\delta Q_{W}=\frac{\Delta Q_{W}}{Q_{W}^{SM}}$. The deviation $\Delta
Q_{W}$ is \cite{cesio} 
\begin{equation}
\Delta Q_{W}=\left[ \left( 1+4\frac{S_{W}^{4}}{1-2S_{W}^{2}}\right) Z-N%
\right] \Delta \rho _{M}+\Delta Q_{W}^{\prime },  \label{dev}
\end{equation}
and $\Delta Q_{W}^{\prime }$ which contains new physics gives

\begin{eqnarray}
\Delta Q_{W}^{\prime } &=&-16\left[ \left( 2Z+N\right) \left( g_{A}^{e}%
\overset{\sim }{g}_{V}^{u}+\overset{\sim }{g}_{A}^{e}g_{V}^{u}\right)
+\left( Z+2N\right) \left( g_{A}^{e}\overset{\sim }{g}_{V}^{d}+\overset{\sim 
}{g}_{A}^{e}g_{V}^{d}\right) \right] S_{\theta }  \notag \\
&&-16\left[ \left( 2Z+N\right) \overset{\sim }{g}_{A}^{e}\overset{\sim }{g}%
_{V}^{u}+\left( Z+2N\right) \overset{\sim }{g}_{A}^{e}\overset{\sim }{g}%
_{V}^{d}\right] \frac{M_{Z_{1}}^{2}}{M_{Z_{2}}^{2}}.  \label{new}
\end{eqnarray}

For Cesium we have $Z=55,$ $N=78$, and for the first term in (\ref{dev}) we
take the value $\left[ \left( 1+4\frac{S_{W}^{4}}{1-2S_{W}^{2}}\right) Z-N%
\right] \Delta \rho _{M}\simeq -0.01$ \cite{cesio}. With the definitions of
the electroweak couplings $\overset{\sim }{g}_{V,A}^{f}$ in eq. (\ref{coup1}%
), we can see that the new physics contributions given by eq. (\ref{new}) is $%
\beta$-dependent, so that the precision measurements are sensitive to the
type of 331 model according to the value of $\beta$. This dependence will
allow us to perform precision adjustments to $\beta$, i.e model adjustment.
We get the same correction for the spectrum $A$ and $B$ due to the fact that
the weak charge depends mostly on the up-down quarks, and $A,B$-cases
maintain the same representation for this family.

With the expressions for the Z-pole observables and the experimental data
shown in table \ref{tab:observables}, we perform a $\chi ^{2}$ fit for each
representation $A,B$ and $C$ at 95\% CL, which will allow us to display the
family dependence in the model. The results are resumed in table \ref{tab:bound-1} and \ref{tab:bound-2}. First of all, we find the best allowed region in the plane $S_{\theta }-\beta $ for
three different values of $M_{Z_{2}}$. The lowest bound of $M_{Z_{2}}$ that
displays an allowed region is about $1200$ GeV, which appears only for the C
assignment such as fig. 1 shows. We can see in the figure that models
with negative values of $\beta $ are excluded, including the usual models
with $\beta =-\sqrt{3},-\frac{1}{\sqrt{3}}$. This non-symmetrical behavior in
the sign of $\beta $ is due to the fact that the vector and axial couplings
in eq. (\ref{coup1}) have a lineal dependence with $\beta $, which causes
different results according to the sign. Figs. 2 and 3 display broader
allowed region for $M_{Z_{2}}=1300$ and $4000$ GeV respectively. Thus, the possible 331-models is highly restricted by low values of $M_{Z_{2}}$ (including the exclusion of the main versions), but if the energy scale increases, new 331 versions are accessible. The models from literature are suitable for high values of $Z_{2}-$mass. We also see that for small $Z_{2}-$mass, the bounds associated to the mixing angle are very small ($\sim10^{-4}$).

On the other hand, we obtain the regions in the plane $M_{Z_{2}}-\beta $ for small values of $S_{\theta}$. Figs. 4 and 5 show regions for negative mixing angle, which favour models with $\beta<0$. It is interesting to note that regions A and B display an upper bound for $M_{Z_{2}}$ when $S_{\theta}=-0.0008$. Figs. 6 and 7 show regions for positive mixing angles. In particular, we can see in fig. 7 that if $S_{\theta}=0.001$, the C-family assignment does not display allowed region. In all plots we note that $A$-region and $B$-region are very similar
because they present the same weak corrections; the small differences arise mostly due to the bottom
correction in eq. (\ref{shift3}). We emphasize that although these results admit continuous values of $\beta$ (including zero), under some circumstances there are additional restrictions from basic principles that could forbid some specific values, as it is studied in ref. \cite{beta-arbitrary}. For instance, the model with $\beta=0$ does not generate spontaneous symmetry breaking, which is required to provide the mass spectrum.

\begin{table}[tbp]
\begin{center}
\begin{tabular}{|c|c|c|c|}
\hline
$M_{Z_{2}}$ (GeV) & Quarks Rep. & $\beta $ & $S_{\theta }$ ($\times 10^{-4}$)
\\ \hline
& Rep. $A$ & No Region & No Region \\ \cline{2-4}
$1200$ & Rep. $B$ & No Region & No Region \\ \cline{2-4}
& Rep. $C$ & $1.1\lesssim \beta \lesssim 1.73$ & $-1\leq S_{\theta }\leq 0.7$
\\ \hline\hline
& Rep. $A$ & $-0.1\lesssim \beta \lesssim 1.55$ & $-3\leq S_{\theta }\leq 2$
\\ \cline{2-4}
1300 & Rep. $B$ & $-0.1\lesssim \beta \lesssim 1.55$ & $-3\leq S_{\theta
}\leq 2$ \\ \cline{2-4}
& Rep. $C$ & $0.85\lesssim \beta \lesssim 1.75$ & $-1\leq S_{\theta }\leq 1$
\\ \hline\hline
& Rep. $A$ & $-1.73\lesssim \beta \lesssim 1.8$ & $-8\leq S_{\theta }\leq 19$
\\ \cline{2-4}
4000 & Rep. $B$ & $-1.73\lesssim \beta \lesssim 1.8$ & $-8\leq S_{\theta
}\leq 19$ \\ \cline{2-4}
& Rep. $C$ & $-1.3\lesssim \beta \lesssim 1.8$ & $-9\leq S_{\theta }\leq 7$
\\ \hline
\end{tabular}%
\end{center}
\caption{\textit{Bounds}\textit{ for} $\beta$\textit{ and S}$_{\protect\theta }$ \textit{for
three quark representations} \textit{at 95\% CL and three} $Z_{2}$\textit{-mass} }
\label{tab:bound-1}
\end{table}

\begin{table}[tbp]
\begin{center}
\begin{tabular}{|c|c|c|c|}
\hline
$S_{\theta }$ ($\times 10^{-4}$) & Quarks Rep. & $\beta $ & $M_{Z_{2}}$ (GeV)
\\ \hline
& Rep. $A$ & $-0.76\lesssim \beta \lesssim -0.24$ & $1500\lesssim
M_{Z_{2}}\lesssim 3800$ \\ \cline{2-4}
$-8$ & Rep. $B$ & $-0.78\lesssim \beta \lesssim -0.22$ & $1500\lesssim
M_{Z_{2}}\lesssim 4200$ \\ \cline{2-4}
& Rep. $C$ & $-0.78\lesssim \beta \lesssim -0.22$ & $2000\lesssim M_{Z_{2}}$
\\ \hline\hline
& Rep. $A$ & $-1.13\lesssim \beta \lesssim 0.3$ & $1400\lesssim M_{Z_{2}}$
\\ \cline{2-4}
$-5$ & Rep. $B$ & $-1.13\lesssim \beta \lesssim 0.3$ & $1400\lesssim
M_{Z_{2}}$ \\ \cline{2-4}
& Rep. $C$ & $-1.13\lesssim \beta \lesssim 0.3$ & $2000\lesssim M_{Z_{2}}$
\\ \hline\hline
& Rep. $A$ & $-1.4\lesssim \beta \lesssim 0.5$ & $1500\lesssim M_{Z_{2}}$ \\ 
\cline{2-4}
5 & Rep. $B$ & $-1.4\lesssim \beta \lesssim 0.5$ & $1500\lesssim M_{Z_{2}}$
\\ \cline{2-4}
& Rep. $C$ & $-1.1\lesssim \beta \lesssim 0.3$ & $2800\lesssim M_{Z_{2}}$ \\ 
\hline\hline
& Rep. $A$ & $-1.05\lesssim \beta \lesssim 0.05$ & $1500\lesssim M_{Z_{2}}$ \\
\cline{2-4}
10 & Rep. $B$ & $-1.05\lesssim \beta \lesssim 0.05$ & $1500\lesssim M_{Z_{2}}
$ \\ \cline{2-4}
& Rep. $C$ & No Region & No Region \\ \hline
\end{tabular}%
\end{center}
\caption{\textit{Bounds}\textit{ for }$\beta$\textit{ and
M}$_{Z_{2}}$  \textit{for three quark
representations} \textit{at 95\% CL and four mixing angle} $S_{\theta}$ }
\label{tab:bound-2}
\end{table}

\section{Conclusions\label{conclusions}}

The $SU(3)_{c}\otimes $ $SU(3)_{L}\otimes U(1)_{X}$ models for three
families with $\beta$ arbitrary was studied under the framework of family dependence.

As it is shown in table \ref{tab:combination}, we found three different
assignments of quarks into the mass family basis. Each assignment determines
different weak couplings of the quarks to the extra neutral current
associated to $Z_{2}$, which holds a small angle mixing with respect to the
SM-neutral current associated to $Z_{1}.$ This mixing gives different
allowed regions in the $S_{\theta }-\beta$ and $M_{Z_{2}}-\beta$ planes for the LEP parameters
at the Z-pole and including data from the atomic parity violation.

Performing a $\chi ^{2}$ fit at 95\% CL we found regions $S_{\theta }-\beta$ that display a dependence in the family assignment for different values of $M_{Z_{2}}$ (figs. $1-3$). For the lowest value $M_{Z_{2}}=	1200$ GeV, we found that only those 331 models with $1.1\lesssim\beta\lesssim1.73$ and quarks families in the C-representation yield a possible region with small mixing angle ($\sim10^{-4}$). The possibilities of 331-models grow as $M_{Z_{2}}$ grows, exhibiting broader regions for the mixing angle.
For the $M_{Z_{2}}-\beta$ plots (figs. $4-7$), we also found model and family restrictions according to the mixing angle. In this case the $\beta$-bound grows when the mixing angle decreases near zero. This behavior seen in the four figures is in agreement with the results from figs. $1-3$, where the bounds for $\beta$ acquire their maximum values around $S_{\theta}=0$. The Pleitez and Long models ($\beta =-\sqrt{3},-\frac{1}{\sqrt{3}}$ respectively) are excluded for low values of $M_{Z_{2}}$ ($\leq1200$ GeV).      
     
Unlike the SM where the family assignment is arbitrary without any
phenomenological change, our results show how this assignment yields
differences in the numerical predictions for 331 models. We see that the
lowest bound for $M_{Z_{2}}$ is higher than those obtained by other authors
for one family models \cite{fourteen}. Due to the restriction of the data
from the atomic parity violation, we are getting a differences of about one
order of magnitude in the lowest bound for the $M_{Z_{2}}$.

This study can be extended if we consider linear combinations among the
three familiy assignments according to the ansatz of the quarks mass matrix
in agreement with the physical mass and mixing angle mass. In this case, the
allowed regions would be a combination
among the regions obtained here.

We acknowledge the financial support from COLCIENCIAS.

\newpage

\begin{figure}[tbph]
\centering \includegraphics[scale=0.9]{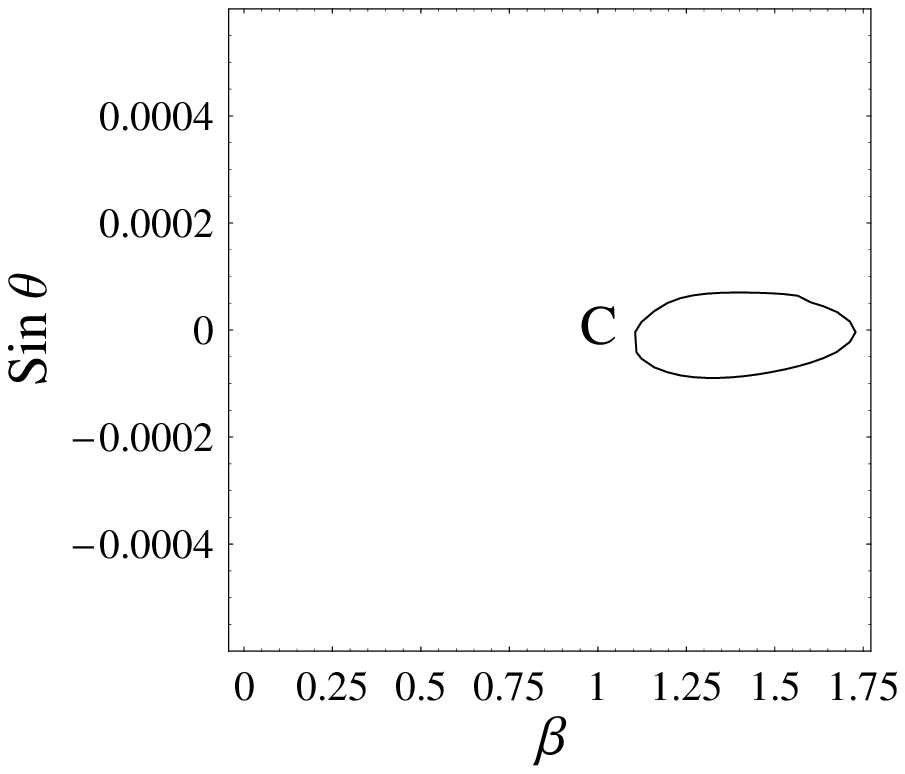}
\caption{\textit{The allowed region for $\sin\protect\theta$ vs $\protect%
\beta$ with $M_{Z_{2}}=1200$ GeV. C correspond to the assignment of family
from table 2. A and B assignments are excluded at this scale.}}
\label{figura1}
\end{figure}

\begin{figure}[tbph]
\centering \includegraphics[scale=0.8]{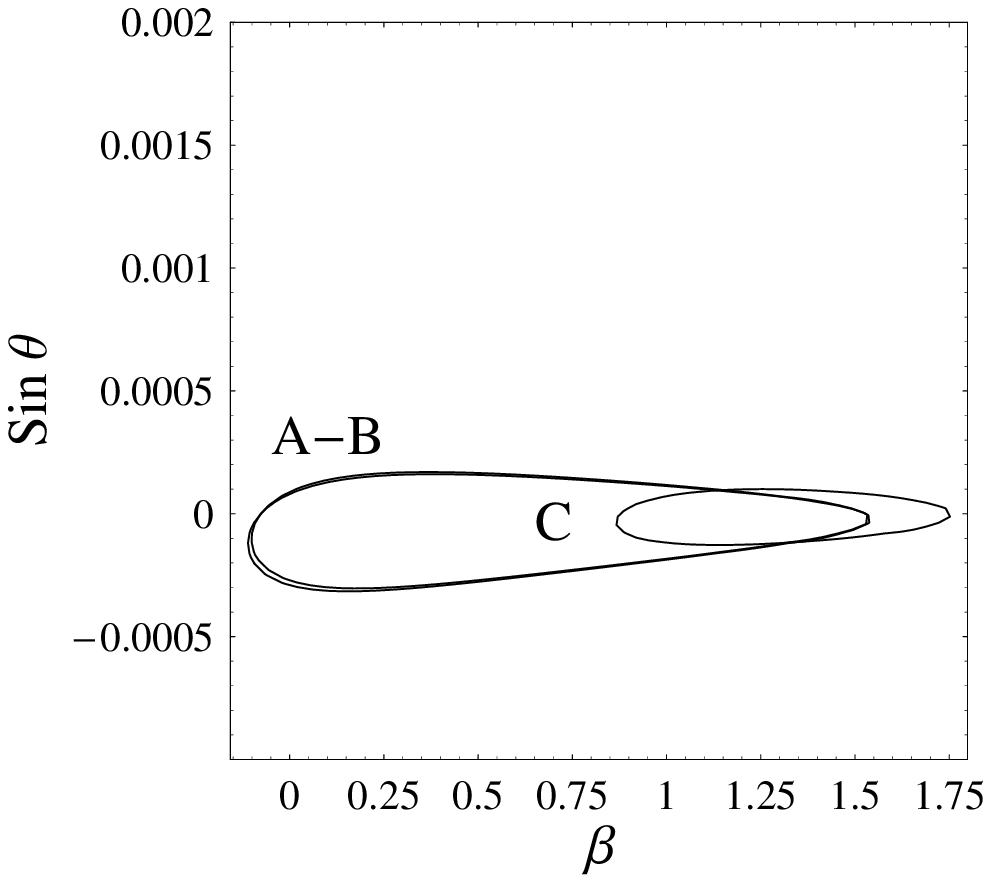}
\caption{\textit{The allowed region for $\sin\protect\theta$ vs $\protect%
\beta$ with $M_{Z_{2}}=1300$ GeV. A, B and C correspond to the assignment of
families from table 2.}}
\label{figura2}
\end{figure}

\begin{figure}[tbph]
\centering \includegraphics[scale=0.8]{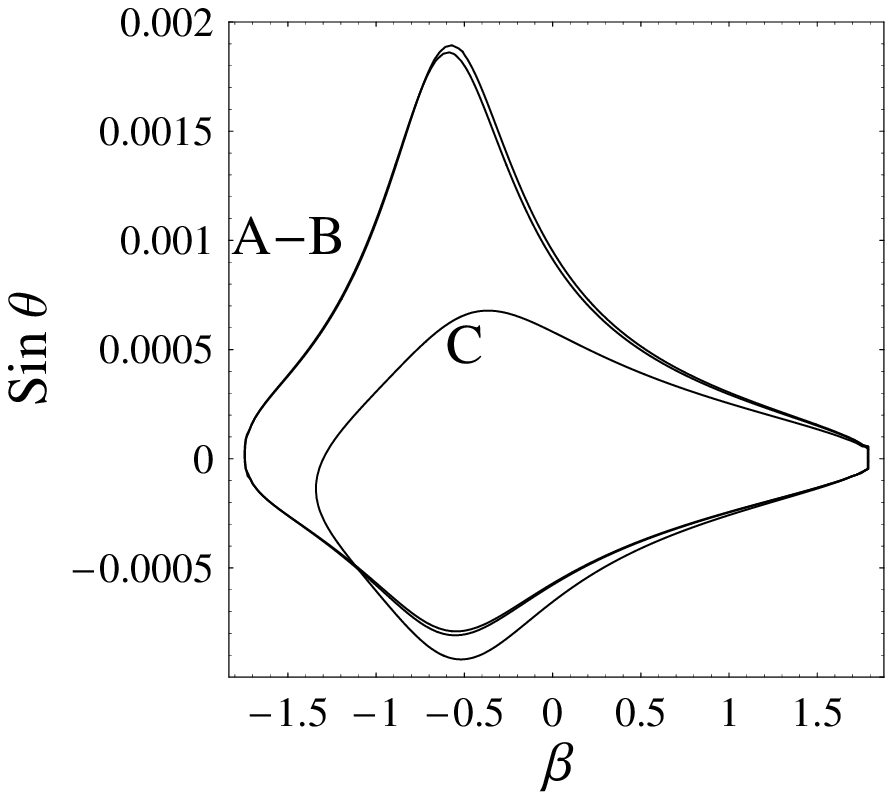}
\caption{\textit{The allowed region for $\sin\protect\theta$ vs $\protect%
\beta$ with $M_{Z_{2}}=4000$ GeV. A, B and C correspond to the assignment of
families from table 2}}
\label{figura3}
\end{figure}

\begin{figure}[tbph]
\centering \includegraphics[scale=0.8]{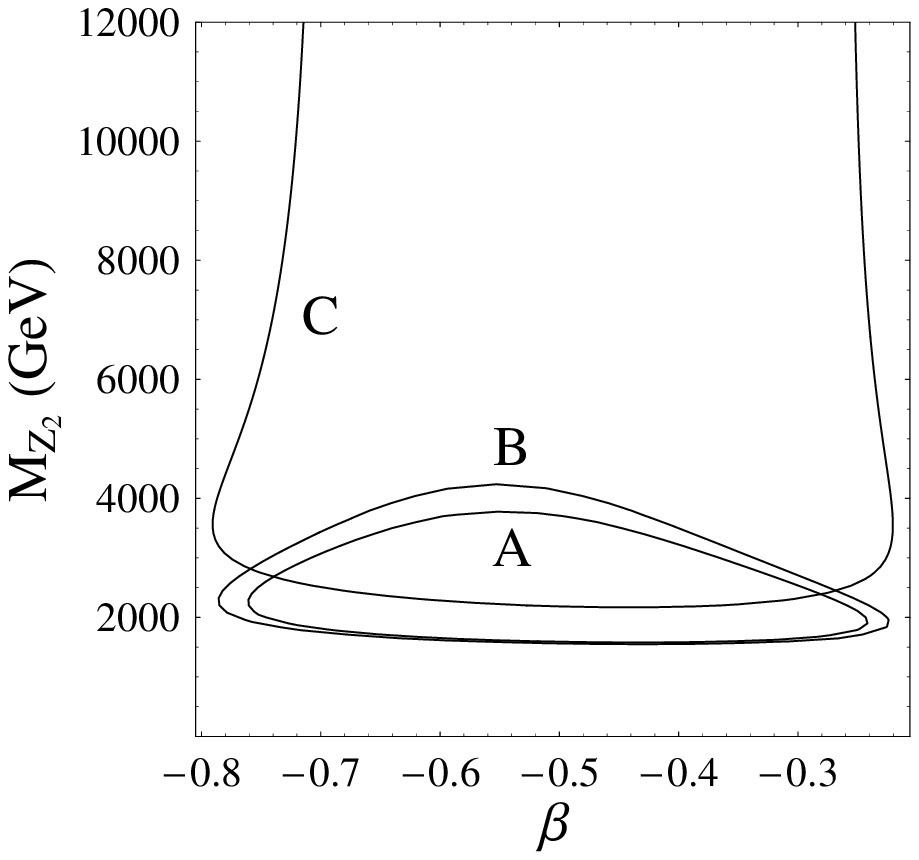}
\caption{\textit{The allowed region for $M_{Z_{2}}$ vs $\protect\beta$ with $%
\sin\protect\theta=-0.0008$. A, B and C correspond to the assignment of
families from table 2}}
\label{figura4}
\end{figure}

\begin{figure}[tbph]
\centering \includegraphics[scale=0.8]{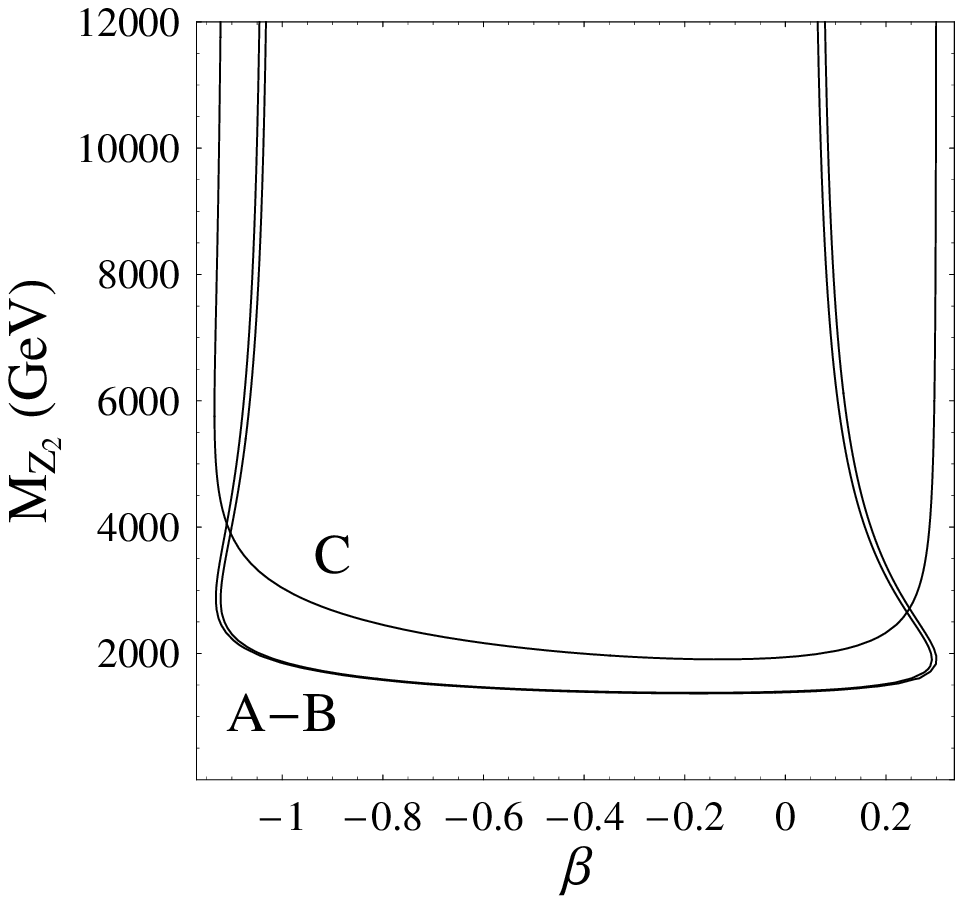}
\caption{\textit{The allowed region for $M_{Z_{2}}$ vs $\protect\beta$ with $%
\sin\protect\theta=-0.0005$. A, B and C correspond to the assignment of
families from table 2}}
\label{figura5}
\end{figure}

\begin{figure}[tbph]
\centering \includegraphics[scale=0.8]{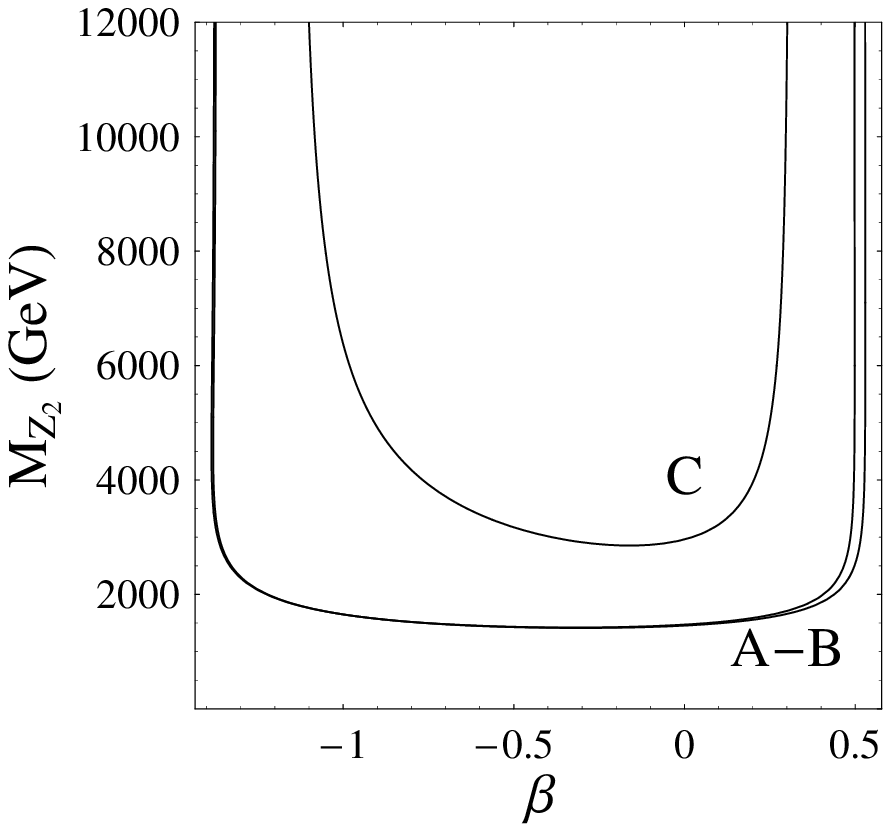}
\caption{\textit{The allowed region for $M_{Z_{2}}$ vs $\protect\beta$ with $%
\sin\protect\theta=0.0005$. A, B and C correspond to the assignment of
families from table 2}}
\label{figura6}
\end{figure}

\begin{figure}[tbph]
\centering \includegraphics[scale=0.8]{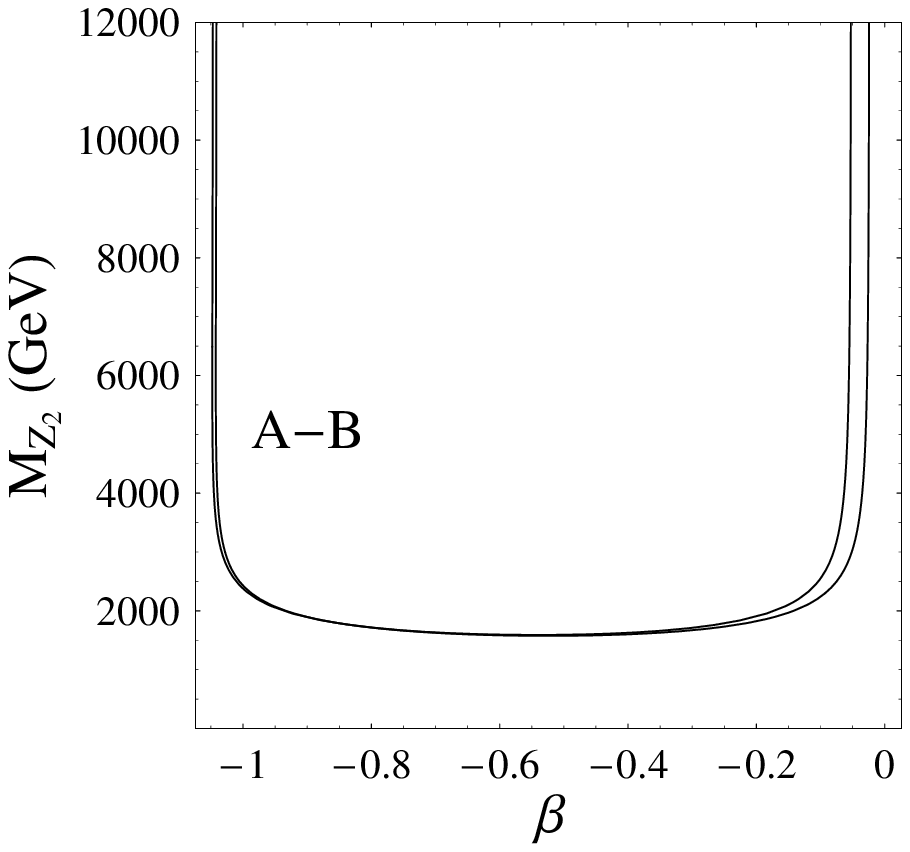}
\caption{\textit{The allowed region for $M_{Z_{2}}$ vs $\protect\beta$ with $%
\sin\protect\theta=0.001$. A and B correspond to the assignment of families
from table 2. C is excluded for this mixing angle.}}
\label{figura7}
\end{figure}


\begin{thebibliography}{99}
\bibitem{zprima} For a review see A. Leike, Phys. Rep. \textbf{317},
143 (1999); J. Erler, P. Langacker, T.J. Li, Phys. Rev. \textbf{D66},
015002 (2002); S. Hesselbach, F. Franke, H. Fraas, Eur. Phys. J. \textbf{C23}%
, 149 (2002).

\bibitem{godfrey} S. Godfrey in Proc. of the APS/DPF/DPB Summer Study on the
Future of Particles Physics (Snowmass 2001), ed. N. Graf, arXiv:
hep-ph/0201093 and hep-ph/0201092; Marcela Carena, Alejandro Daleo, Bogdan
A. Dobrescu, Tim M.P. Tait, Phys. Rev. \textbf{D70}, 093009 (2004).

\bibitem{Frampton2} For a review about the family problem see P.H. Frampton, P.Q. Hung and M. Sher, arXiv:
hep-ph/9903387 v2.

\bibitem{anomalias} J.S. Bell, R. Jackiw, Nuovo Cim. \textbf{A60} 47 (1969);
S.L. Adler, Phys. Rev. \textbf{177}, 2426 (1969); D.J. Gross, R. Jackiw,
Phys.Rev. \textbf{D6} 477 (1972). H. Georgi and S. L. Glashow, Phys. Rev. 
\textbf{D6} 429 (1972); S. Okubo, Phys. Rev. \textbf{D16}, 3528 (1977); J.
Banks and H. Georgi, Phys. Rev. \textbf{14} 1159 (1976).

\bibitem{Pires} C.A. de S. Pires and O.P. Ravinez, Phys. Rev. \textbf{D58},
35008 (1998); C.A. de S. Pires , Phys. Rev. \textbf{D60}, 075013 (1999).

\bibitem{fourteen} L.A. S\'{a}nchez, W.A. Ponce, and R. Mart\'{\i}nez, Phys.
Rev. \textbf{D64}, 075013 (2001); R. Martinez, William A. Ponce, Luis A.
Sanchez, Phys. Rev.\textbf{D65} 055013 (2002); W.A. Ponce, J.B. Fl\'{o}rez
and L.A. S\'{a}nchez, Int. J. Mod. Phys. \textbf{A17}, 643 (2002); W.A.
Ponce, Y. Giraldo, and L.A. S\'{a}nchez, Phys. Rev. \textbf{D67}, 075001
(2003).

\bibitem{ten} F. Pisano and V. Pleitez, Phys. Rev. \textbf{D46}, 410 (1992);
P.H. Frampton, Phys. Rev. Lett. \textbf{69}, 2889 (1992); R. Foot, O.F.
Hernandez, F. Pisano and V. Pleitez, Phys. Rev. \textbf{D47}, 4158 (1993);
P.H. Frampton, P. Krastev and J.T. Liu, Mod. Phys. Lett. \textbf{9A}, 761
(1994); P.H. Frampton et. al. Mod. Phys. Lett. \textbf{9A}, 1975 (1994);
Nguyen Tuan Anh, Nguyen Anh Ky, Hoang Ngoc Long, Int. J. Mod. Phys. \textbf{%
A16}, 541 (2001); J.C. Montero, C.A.deS. Pires and V. Pleitez, Phys. Rev.
D65 095001 (2002). 

\bibitem{twelve} R. Foot, H.N. Long and T.A. Tran, Phys. Rev. \textbf{D50},
R34 (1994); H.N. Long, Phys. Rev. \textbf{D53}, 437 (1996); \textit{ibid}, 
\textbf{D54}, 4691 (1996); H.N. Long, Mod. Phys. Lett. \textbf{A13}, 1865
(1998). 

\bibitem{331us} Rodolfo A. Diaz, R. Martinez, F. Ochoa, Phys. Rev. \textbf{\
D69}, 095009 (2004).

\bibitem{beta-arbitrary} Rodolfo A. Diaz, R. Martinez, F. Ochoa, arXiv:
hep-ph/0411263 (will be published in Phys. Rev. D).

\bibitem{Frampton} P.H. Frampton and D. Ng, Phys. Rev. \textbf{D45}, 4240
(1992).

\bibitem{family-dependence} Fredy Ochoa and R. Martinez, arXiv: hep-ph/0505027
(will be published in Phys. Rev. D).

\bibitem{Mohapatra} In this connection see K.T. Mahanthappa and P.K.
Mohapatra, Phys. Rev. \textbf{D42}, 1732-2400 (1990); Phys. Rev. \textbf{D43}%
, 3093 (1991).

\bibitem{one} S. Eidelman et. al. Particle Data Group, Phys. Lett. \textbf{%
B592}, (2004), pp. 120-121.

\bibitem{pitch} J. Bernabeu, A. Pich and A. Santamaria, Nucl. Phys. \textbf{%
B363} 326 (1991); D. Bardin et.al. Electroweak Working Group Report, arXiv:
hep-ph/9709229 (1997) 28-32.

\bibitem{greub} F.M. Borzumati and C. Greub, Phys. Rev. D\textbf{58}, 074004
(1998).

\bibitem{cesio} G. Altarelli, R. Casalbouni, S. De Curtis, N. Di Bartolomeo,
F. Feruglio and R. Gatto, Phys. Lett. \textbf{B 261} 146 (1991); H.N. Long
and L.P. Trung, Phys. Lett. \textbf{B502} (2001) 63-68.
\end{thebibliography}
\end{document}